\documentclass[aps,prd,12pt,
               preprintnumbers,numbers,sort&compress,nofootinbib,
               showpacs]{revtex4}
\usepackage{amsmath}
\usepackage{amsmath}
\usepackage{amssymb}
\usepackage{psfrag}
\usepackage{epsfig}
\usepackage{graphicx}
\usepackage{epstopdf}
\newcommand{\exclude}[1]{}

\begin{document}
\newcommand{\be}{\begin{equation}}
\newcommand{\ee}{\end{equation}}
\def\Journal#1#2#3#4{{#1} {\bf #2}, #3 (#4)}

\newcommand{\A}{\alpha}
\newcommand{\B}{\beta}
\newcommand{\T}{\theta}
\newcommand{\Ep}{\epsilon}
\newcommand{\beq}{\begin{equation}}
\newcommand{\eeq}{\end{equation}}
\newcommand{\fr}{\frac}
\newcommand{\beqn}{\begin{eqnarray}}
\newcommand{\eeqn}{\end{eqnarray}}
\newcommand{\G}{\gamma}
\newcommand{\D}{\delta}
\renewcommand{\P}{\phi}
\newcommand{\intl}{\int\limits_{1}^{\infty}}
\renewcommand {\L}{\lambda}
\newcommand{\pt}{\partial}

\newcommand{\bq}{\bar q_A}
\newcommand{\tq}{\tilde q_A}
\newcommand{\btq}{\bar{\tilde q}^A}
\newcommand{\fa}{\varphi^A}
\newcommand{\bfa}{\bar \varphi_A}

\newcommand{\none}{{\cal N}=1}                            
\newcommand{\ntwo}{{\cal N}=2}                            

\def\st{\scriptstyle}
\def\sst{\scriptscriptstyle}
\def\mco{\multicolumn}
\def\epp{\epsilon^{\prime}}
\def\vep{\varepsilon}
\def\ra{\rightarrow}
\def\al{\alpha}
\def\ab{\bar{\alpha}}
\def\bea{\begin{eqnarray}}
\def\eea{\end{eqnarray}}

\begin{flushright}
ITEP-TH-03/09
\end{flushright}

\title{  On Classification of QCD defects via holography}

\author{   Alexander S. Gorsky, Valentin I. Zakharov}
\affiliation{Institute of Theoretical and Experimental
Physics, B. Cheremushkinskaya ul. 25, 117259
Moscow, Russia}
\author{Ariel R. Zhitnitsky}

\affiliation{Department of Physics and Astronomy, University of
  British Columbia,  Vancouver,  BC  V6T1Z1,Canada}

\begin {abstract}
We discuss classification of defects of various codimensions
within a holographic model of pure Yang-Mills theories
or gauge theories with fundamental matter. We focus on their role
below and above the phase transition point as well as their weights in the partition function.
The general result  is that
objects which are stable and heavy in one phase are becoming very
light (tensionless)  in the other phase. We argue that the $\theta$
dependence of the partition function drastically changes at the phase transition point,
and therefore it correlates
with  stability properties of configurations.
Some possible applications
for study  the QCD vacuum properties  above and below phase transition
are also discussed.
\end{abstract}

\maketitle
\section{Introduction}
The gauge/string duality proved to be effective in description
of various aspects of gauge theories at weak and strong coupling.
In the  N=4 SYM one can discuss a precise comparison of the gauge
theory results with the sigma model or SUGRA calculation since the relevant
geometry $AdS_5\times S^5$ is well established. However the situation in the
theory with less amount of SUSY is much more complicated and the explicit
background for the pure YM or QCD is not found yet. The most useful
dual model of pure YM at nonzero temperature  \cite{wittenterm} is based on a stack of $N_c$ D4 branes
which are wrapped around a compact coordinate and at large $N_c$
provide the geometry of the black hole in $AdS_5$. Adding the probe $N_f$
$D8-\bar{D8}$ branes one obtains the dual picture for
QCD \cite{ss} which reproduces the chiral Lagrangian  and captures
many qualitative aspects of the strong coupling physics.

In this Letter we shall discuss extended objects in
QCD from the dual perspective. In the dual model the natural objects to consider
are probe D-branes. The Sakai-Ssugimoto model is
based on the IIA side of the string theory hence there are stable
D0,D2,D4,D6,D8 D branes as well as NS5 brane which we shall discuss on. We shall try to
get  qualitative picture concerning the role of the
various defects which is insensitive to the details
of the metric.
The background involves two periodic coordinates, angular
coordinate on the black hole cigar $x_4$ and the Euclidean time coordinate $\tau$ which branes can wrap around.

In the previous studies some identifications of the defects have been made.
The D0 brane extended along $x_4$ was identified as the YM instanton
\cite{Bergman:2006xn}. The D2 brane wrapped around both periodic
coordinates was identified as the magnetic string \cite{gz}. The D6 brane wrapped
around the compact $S^4$ part of the dual geometry and $\tau$ was considered
in pure YM \cite{wittenflux} and dual QCD \cite{Hong:2008nh,Lin:2008vv} where it has
interpretation of the domain wall separating two vacua. The D4 brane
wrapped around the $S^4$ and extended along $\tau$ has the interpretation
of the "baryonic vertex" in pure YM and  in the Sakai-Sugimoto model
\cite{ss}. Some interplay between the D branes and the $Z_N$ domain walls
has been discussed in the holographic picture in \cite{wittenah,armoni,yee}.

There are a few generic facts concerning the branes and their intersections. Let us numerate   few of them relevant for  the main text:
\begin{itemize}
\item{ p-brane behaves as an instanton on the (p+4)-brane worldvolume \cite{douglas}}
\item{p-brane parallel to the (p+2)-brane gets melted into  homogenious field \cite{gava}}
\item{p-brane transverse to the {p+2}-brane behaves as the monopole
on the (p+2) worldvolume} \cite{dia}
\item{branes in external fields can expand into higher
dimensional branes via the Myers effect \cite{myers}}
\end{itemize}

Our goal here is to look at a variety of
defects using a universal classification scheme. We will be mostly interested in
behavior of the corresponding configurations when the confinement-deconfinement phase transition
is crossed and shall emphasize  some universal properties of the D-defects.
It is very likely that some of the defects to be discussed here
are very important for physics, some of them could be  irrelevant. Therefore, we anticipate that
some important/interesting  defects will be further discussed and studied  in great detail in future.
It is not the goal of the present paper to go into a deep detail analysis of each particular configuration.
Instead, our goal is the classification of  the D-defects, with emphasis
on the change of their properties  across the phase transition..

To be more specific, our basic tool is the dual description of the deconfinement phase transition as
 the Hawking-Page phase transition \cite{wittenterm}, in which case   the two metrics with the same asymptotic get interchanged.
The wrapping around
$x_4$ is stable above the phase transition $T>T_c$  while it is    unstable below the critical temperature
$T<T_c$, see definitions and details below. And vice-verse, the wrapping around $\tau$ is stable at small
temperatures $T<T_c$ and unstable at high temperatures, $T>T_c$.

Since the nonperturbative physics is sensitive to the $\theta$-term we shall also
discuss the $\theta$ dependence  from the dual perspective. We argue that the behavior
of the D defects  is strongly correlated  with
 the  $\theta$-dependence when the phase transition is crossed at $T_c$.
 Such a drastic change in $\theta$ dependence at the $T_c$
 has already been noticed in the literature   \cite{Toublan:2005tn,Bergman:2006xn, gz, Parnachev:2008fy,Zhitnitsky:2008ha}.
 We note also that some drastic  changes in $\theta$ behavior are also supported by
the numerical lattice results \cite{Alles:1996nm} -\cite{ Lucini:2005vg},
see also a review article \cite{Vicari:2008jw},  which unambiguously
suggest that the topological fluctuations (related to $\theta$ behavior)
are strongly suppressed in deconfined phase, and this suppression becomes more severe with increasing $N_c$.
 Here we shall
present some additional  examples which support this picture.
\exclude{One more topological argument concerns the baryonic charges of the configuration.
We shall argue that branes of higher dimensions provide the possibility for the exotic states with the baryonic charge to exist. }

The paper is organized as follows. In Section II we describe
our model based on the $N_c$ D4 branes with one compact worldvolume coordinate.
In our main Section III   we classify D- defects treated in the probe approximation
when they do not deform the dual geometry.  In Section IV with discuss some composite objects which combine different types of D branes.
Finally , in Section V we introduce matter fields in our system
which treated as the probes, $N_f\ll N_c$. Section VI is our  conclusion.

 \section{Description of the model}
A natural starting point to discuss the dual holographic geometry is
provided by a set of $N_c$ D4 branes wrapped around a compact
dimension \cite{wittenterm}. We shall consider the pure
gauge sector first and then add flavor D8 branes along the
lines of the Sakai-Sugimoto model \cite{ss}.

We shall assume the large $N_c$ limit and consider the supergravity
approximation.
In this approximation the geometry looks as $M_{10}=R_{3,1}\times D \times S^4$ and
the corresponding metric reads as
$$
ds^2=(\frac{u}{R_0})^{3/2}(-dt^2 + \delta_{ij}dx^i dx^j +f(u)dx_4^2)+
(\frac{u}{R_0})^{-3/2}(\frac{du^2}{f(u)} +u^2 d\Omega_{4}^2)
$$
\begin{equation}
e^{\Phi}=(\frac{u}{R_0})^4, \qquad F_4=\frac{3N_c\epsilon_4}{4\pi},\qquad
f(u)=1- (\frac{u_{\Lambda}}{u})^3
\end{equation}
where $R_0=(\pi g_s N_c)^{1/3}$
and
$R=\frac{4\pi}{3}(\frac{R_0^3}{u_{\Lambda}})^{1/2}$.
The coupling constant of Yang-Mills theory is related to the radius
of the compact dimension $R$ as follows
$$
g_{YM}^2=\frac{8\pi ^2 g_s l_s}{R}
$$

At zero temperature theory is in the confinement phase
and in the ($u,x_4$) coordinates we have the geometry
of  a cigar with the tip at $u=u_{\Lambda}$. The D4 branes
are located along our (3+1) geometry and are extended
along the internal $x_4$ coordinate.
The key point is that in the non-zero temperature
case there are two backgrounds with similar asymptotic topology of
$R^3\times S^1_{\tau} \times S^1 \times S^4$, where
$\tau$ is the Wick-rotated time coordinate $\tau=it$,
$\tau\propto \tau + \beta$.
One background corresponds to the  analytic continuation
of the  metric described above while the second background corresponds
to  interchange of $\tau$ and $x_4$, that is
the warped factor is attached to the $\tau$ coordinate
and the cigar geometry emerges in the ($\tau,u$) plane instead
of ($x_4,u$) plane which now exhibits the cylinder geometry, see Figure 1. It was
shown in
\cite{wittenterm} by calculation of the free energies that above $T_c$ the
latter background dominates.

\begin{figure}
\epsfxsize=15cm
\centerline{\epsfbox{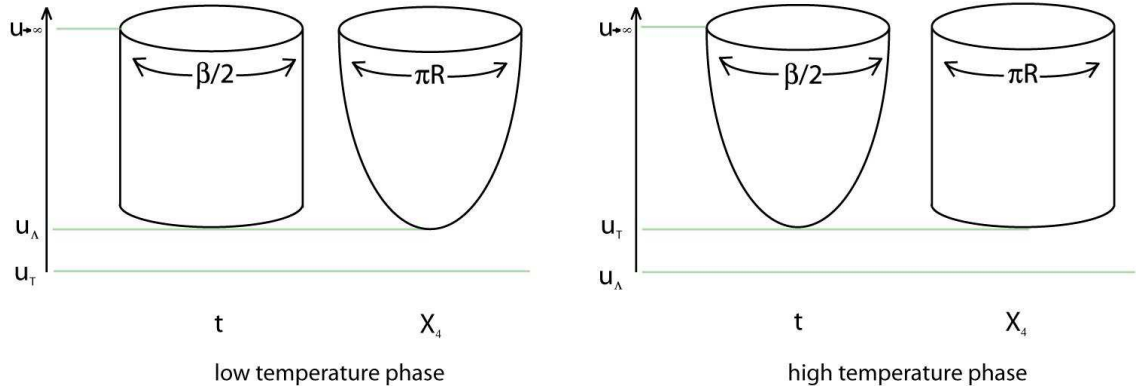}}
\caption{\label{fig1}
  Metric at $T<T_c$ (left figure) and $T>T_c$ (right figure). }
\end{figure}

That is above  phase transition
wrapping around the internal $x_4$ circle is topologically stable,
while the wrapping around the  Euclidean time coordinate
is unstable. This is opposite to the stability pattern of the two wrappings
below the phase transition.

Another issue which we shall be interested in concerns the $\theta$-dependence
of the worldvolume theories on the different probe branes. It can be traced
from the Chern-Simons(CS) terms involving the interaction with the RR one-form $C_1$
\begin{equation}
\delta L= \int C_1 \wedge e^{\cal{F}}
\end{equation}
where $\cal{F}$ is the gauge two-form.
Taking into account that
\begin{equation}
\theta = \int dx_4\ C_1
\end{equation}
one immediately recognizes that the $\theta$ dependence of the worldvolume
theories on the defects correlates with the wrapping around $x_4$ coordinate.
Moreover it is clear that the $\theta$ dependence
of a single defect is poorly defined in the
confined phase since the wrapping around $x_4$ is topologically unstable
and one could discuss the $\theta$ dependence of a kind of a condensate of the  defects.

To model QCD one adds the $N_f$ $D8-\bar{D8}$ pairs
localized at points in $x_4$ coordinate. There are a few qualitative phenomena
to be mentioned. First, the chiral symmetry breaking is described geometrically
in terms of the connectness of the  $D8-\bar{D8}$ pair. It was argued that the
restoration of the chiral symmetry and the deconfinement phase transition
generically take place at different temperatures \cite{kutasov1}. Another
essential point concerns the baryonic degrees of freedom. Geometrically baryons
are identified as the D4 branes wrapped around  the compact $S^4$ and they
are the instantons in the D8 brane worldvolume theory,
Note that in the holographic QCD there should be some care concerning the gauge invariance
of the RR field.
The gauge invariant
field strength of the RR field  due to the bulk anomaly  gets shifted
by the $\eta'$ meson and the correct invariant
identification of the $\theta$-term reads as
\beq
\int_{D} F_{2,inv}=\theta + \frac{\sqrt{2N_f} }{f_{\pi}}\eta'
\eeq
where the integration over the ${u,x_4}$ disc is implied.

In the thermal gauge theory a natural order parameter  is the vacuum expectation value of the Polyakov loop
\begin{equation}
<W(\beta)>= <Tr P exp(\int d\tau A_0)>
\end{equation}
which is vanishing at $T<T_c$ while $<W(\beta)>\neq 0$ at $T>T_c$. This implies
that $Z_N$ symmetry is unbroken at   $T<T_c$ and broken at $T>T_c$. It is possible
to discuss another R-type symmetry of the rotation of $x_4$ coordinate
which is assumed to be broken nonperturbatively at zero
temperature  to the discrete one  similar to SUSY case.
Therefore one could expect the total discrete symmetry
to be $Z_N\times Z_N$ where the order parameter for the second factor is
\begin{equation}
\label{W(R)}
 <W(R)>= <Tr P exp(\int d x_4 A_4)>
\end{equation}
It can serve as the order parameter analogous to the Polyakov loop since
it has a nontrivial  vev at small temperatures and vanishes in the deconfinement phase.
Let us emphasize that the total discrete group mentioned above differs from the same
product discussed in \cite{wittenah}. In that paper the second factor corresponds
to the S-dual magnetic center group whose order parameter is identified with
the T'Hooft loop.

\section{Zoo of the D- defects}
\subsection{D0 branes}
{\it \underline{D0  instantons}.}\\
The simplest defects to be discussed are D0 branes.
It was argued in \cite{Bergman:2006xn} that  instantons
are represented by the Euclidean D0 branes wrapped around
$x_4$. In that case it was argued that the instanton is well defined
above $T_c$ as it corresponds to  the the geometry of the cylinder.
On the other hand, any finite number of instantons
are ill-defined below $T_c$ because of the D0-brane instability.

The $\theta$ dependence
of the D0 action is captured by the CS term on its
worldline. The change of the
instanton role at the transition point corresponds
to the change from the Witten-Veneziano to t'Hooft
mechanisms of the solution to the $U(1)$ problem.
In QCD-like brane setup D0 branes wrapped around $x_4$
intersect with the flavor branes and induce the sources
on the flavor brane worldvolumes.

This picture can also  be readily understood in the quantum-field theory terms since
an estimation for $T_c$ in the $\Lambda_{QCD}$ units   can be given
\cite{Parnachev:2008fy}. Indeed,
the wrapping around $x_4$  corresponds to the well
 defined small instanton and one can use the standard instanton calculus to estimate
 the critical temperature $T_c$ and the $\theta$ behavior above $T_c$:
  \beqn
   \label{T}
   V_{\rm inst}(\theta) &\sim& \cos\theta \cdot e^{-\alpha N\left(\frac{T-T_c}{T_c}\right)}, ~~~~ 1\gg \left(\frac{T-T_c}{T_c}\right)\gg 1/N,
    \nonumber \\
    \chi (T) &\sim& \frac{\partial^2 V_{\rm inst}(\theta)}{\partial\theta^2}\sim
      e^{-\alpha N\left(\frac{T-T_c}{T_c}\right)}\rightarrow 0,  ~~~ \alpha \sim 1, ~~~ N\gg 1.
   \eeqn
  Such a behavior    implies that the dilute gas approximation at large $N_c$ is
justified even in close vicinity of $T_c$ as long as $\frac{T-T_c}{T_c}\gg \frac{1}{N_c}$.
  Such a sharp behavior of the topological susceptibility $ \chi (T)$
is supported by  numerical lattice results \cite{Alles:1996nm} -\cite{ Lucini:2005vg} which unambiguously
suggest that the topological fluctuations are strongly suppressed in the deconfined phase,
and this suppression becomes more severe with increasing $N_c$. These general features observed in the lattice simulations
have very simple explanation within  QFT framework as eq.  (\ref{T}) shows,  as well
as in holographic model of QCD \cite{Bergman:2006xn,Parnachev:2008fy}.

Finally, let us address the following question:  what happens with our D0 instantons
in the deconfined phase,   immediately at $T>T_c$? We know that at sufficiently large temperatures
$\left({T-T_c}\right)/{T_c}\gg 1/N$
the configuration  becomes  a  stable instanton   in 4d  with the size $\rho\sim {(\pi T)}^{-1}$. The density of the instantons
is exponentially suppressed  $\sim \cos\theta \cdot e^{-\alpha N\left(\frac{T-T_c}{T_c}\right)}$,
magnetic charges of the constituents (if exist, see section IV) are completely screened
such that it makes no sense to speak about individual constituents in this regime.
However, for finite $N$ there is a window of temperatures $0<\left({T-T_c}\right)/{T_c}\leq 1/N$
when the magnetic degrees of freedom are not completely screened yet. This window which shrinks to
a point at $N=\infty$  is  obviously beyond  analytical control.
However, these magnetic degrees of freedom  could be be extremely important in the
 window $0<\left({T-T_c}\right)/{T_c}\leq 1/N$. It is tempting to assume that these magnetic degrees
of freedom is a trace of fractional instanton constituents which likely to exist in confined phase, see discussions   in section IV.\\

{\it \underline{D0 - particle}}\\
The  orientation of the D0 brane worldline along the
Euclidean time $\tau$ corresponds to its realization
as a KK particle without the $\theta$ dependence.
However the Polyakov loop $Tr P\exp(\int d\tau A_(\tau))$
may  develop. This configuration at $T<T_c$  is a stable scalar
glue-like configuration which must have very different properties
in comparison with all  standard glueballs when the temperature approaches $T_c$  from below,
$(T_c-T) \rightarrow 0$.   Above the critical temperature (deconfinement phase)
KK modes tend to condense near the tip of the cigar
because of the instability of the wrapping around $ \tau$.
In the deconfinement phase KK modes behave as the instanton -like configuration (with no $\theta$ dependence)
in the effective 3D gauge theory.
This instability  may have enormous consequences for physics
since an arbitrary large number of such states can be produced in vicinity of $T\simeq T_c$.
We shall not elaborate on this issue in the present work.

\subsection{D2 branes}
{\it \underline{D2 string}.}\\
The magnetic string
is the probe D2 brane wrapped
around $S_1$ parameterized by $x_4$ and its tension
is therefore proportional to the effective radius $R(u)$
\cite{gz}.
At small temperatures   this wrapping
is topologically unstable and the D2 brane tends to shrink to the tip
where its tension vanishes. This is the large-$N_c$
counterpart of the  effect of dissolving of $p$-branes inside
$p+2$-branes \cite{gava}.
We see that in this way one immediately
reproduces the observed  property of tensionlessness  of the magnetic string in the
confining phase which however becomes tensionful above the critical temperature $T_c$ of the
deconfinement phase transition.

The  $\theta$-term in the magnetic string
worldsheet  Lagrangian is induced by CS term
\begin{equation}
L_{CS}=\int d^3 x\ C_1\wedge F
\end{equation}
that is configurations with the flux on the worldsheet
amount to the nontrivial 4d topological charge.
It was also argued that the magnetic strings amount to the negative contribution to the total energy of plasma
at $T>T_c$ \cite{gz}.  It could explain the negative-sign
contribution of the lattice magnetic strings into the energy of
plasma \cite{sign}. Because of the instability of the "thermal"
cigar magnetic string becomes effectively particle-like object
in the Euclidean 3D \cite{gz}, in agreement with the lattice studies
\cite{langfeld}.\\

{\it \underline{D2 domain walls}.}\\
Turn now to the discussion of the D2 domain walls. Let us emphasize from the
very beginning that to consider the stable infinite domain walls the
degenerate vacua should exist. On the other hand we expect the single
stable vacuum in the pure YM case and QCD. That is the arguments concerning
the domain walls should be interesting in two aspects. First, there are
metastable vacua whose energy density differs from the density of the
true one  by the terms $O(1/N)$ that is they are almost stable at large N.
One can also consider the domain wall balls when the configuration
is stabilized by the domain wall tension.

There are two types of domain walls in $R^3$  built from D2 branes.
Consider first a D2 brane localized in the $x_4$ coordinate. It corresponds
to a domain wall in 4D and has no $\theta$ dependence. The theory on the domain
wall involves the periodic real scalar field which corresponds to the position
of the D2 brane on the $x_4$ circle as well as a scalar corresponding to its
radial coordinate.
Its tension is small in the deconfinement phase and it behaves as the
string in the 3d effective description. It has no $\theta$ dependence.

The second type of D2 S-domain walls extended in $x_4$  involves
a nontrivial $\theta$ dependent term. Because of the unbroken
electric $Z_N$ symmetry in the confinement phase such domain
walls are expected to exist in N-tuples in this phase
symmetrically on the $\tau$ circle. The worldsheet theory in the
deconfinement phase involves only one real scalar corresponding to
its radial coordinate. These D2 domain walls may play an important dynamical  role
supporting the constituents with fractional topological charges, see Section IV.
\\

{\it \underline{Space filling D2 brane }.}\\
One could also  consider the D2 branes localized both in $\tau$ and $x_4$ directions.
Such space-filling D2 branes most probably are expected to exist in N-tuples
in both phases because in each phase  the D2 brane has  one unbroken $Z_N$
symmetry.
Hence the effective gauge theory  should have  SU(N)
gauge theory in both phases. These branes could play an important
role in the effective 3D description of the  deconfinement phase.
Their worldvolume theory naturally involves one complex and one real
periodic scalars.\\

{\it \underline{ D2-$\bar{D2}$ pair }.}\\
One could also discuss the D2 brane extended along the radial coordinate. Such a configuration
is an analogue of the D8 brane in Sakai-Sugimoto model which is extended
along the radial coordinate and has U-shape form in the chirally   broken phase.
In the holographic QCD case  one actually has $D8-\bar{D8}$ pair which connectness indicates
non-vanishing of the chiral condensate.
In the case of the U-shaped D2 brane we have no flavor brane hence
the chirality can not be defined in the conventional way. However, there is a configuration
which is readily prepared to provide the chiral condensate if D8 brane is introduced into the system.
  If such U-shaped D2 brane is extended along
$\tau$ it behaves as string while in the opposite case as the domain wall.
In both cases the tension of the object is finite. It is tempting  to speculate
that some  kind of the chiral symmetry breaking happens
in pure gauge theory  (without flavor fermions)  being localized at lower dimensional
defects rather than  in the entire  space.

It is also tempting to speculate that some of the D2 branes discussed in this subsection
could be mapped into  the low-dimensional, chirality-related structures  observed on the latices.

\subsection{D4 branes}

{\it \underline{D4 particle}}\\
There are several possible embeddings of D4 branes.  One possibility
concerns  D4 wrapped around $S^4$ and extended along the Euclidean time $\tau$.
The familiar example
of such wrapping in the QCD like geometry \cite{ss} has the interpretation of baryon
if matter fields in the form of D8 branes are present in the system.
The key point here is that due to the CS term
\begin{equation}
\label{D-particle}
\int  d^5 x C_3\wedge F \sim N_c
\end{equation}
the ``electric charge" $N_c$ is induced
on the  D4 brane that is one has to add $N_c$ open strings.
 It is a static topologically stable  configuration.
In the QCD-like case these open strings end on the D8 brane
yielding the baryonic state.

In the pure YM case there are no flavor branes that is one  has to add additional
low-dimensional branes to compensate charge and make a gauge invariant object.
 The most simple way to achieve this  goal is to add
$N_c$ D0 branes yielding the D0-D4 open strings. Hence we get the D4 particle
which does not feel the $\theta$-term and is well-defined below the critical
temperature $T<T_c$.
 \exclude{Due to the unbroken magnetic symmetry in the deconfinement phase D4 particles are expected
to exist in N-tuples.}

\begin{figure}
\epsfxsize=10cm
\centerline{\epsfbox{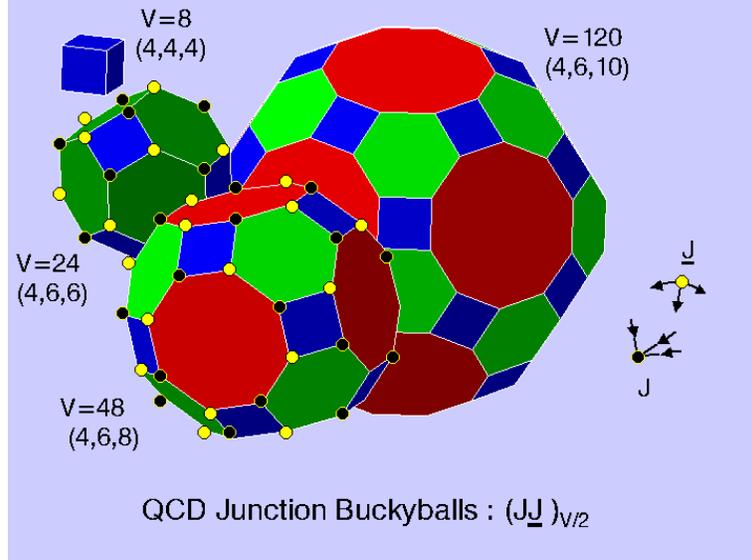}}
\caption{\label{fig2}
 Buckyballs with $N=3$ and $V=8, 24, 48, 120$ from \cite{Csorgo:2001sq}.
The construction combines  `` D4 particle"
with  ``D4 anti-particle" to form a gauge invariant object with zero baryon charge. The mass of this object scales as $\sim N_c$. }
\end{figure}

We can  combine the `` D4 particle"
with  ``D4 anti-particle" to form a gauge invariant object, see FIG 2.
The mass of this object scales as $\sim N_c$ and is much heavier  than the usual glueballs. In fact, one can construct
gauge invariant objects with any even number of vertexes such that the total charge vanishes.
It is a new family of glueballs with mass $\sim N_c V$, where $V$ is the number of D4 and anti-D4 particles which form a desired configuration.
It is amusing that such kind of structure in QCD had been previously discussed \cite{Csorgo:2001sq}
motivated by the discovery of the carbonic Fullerenes $C_{60}$ and $C_{70}$ in 1985,
which are nano-scale objects \cite{fullerenes}.
The QCD objects, similar to the carbonic Fullerenes with femto-meter scale were named Buckyballs.
It has been also demonstrated that the ``magic" numbers for  Buckyballs are $V=8, 24, 48, 120$
which correspond to the most symmetric, and likely, most stable configurations \cite{Csorgo:2001sq}.
The properties of this configuration are not sensitive to $\theta$.
The possibility to discover such kind of configurations at RHIC was discussed in
\cite{Csorgo:2001sq}.
\\

{\it \underline{D4 instanton}.}\\
 If we consider wrapped D4 brane
extended along $x_4$ we get a new-object, "D4 instanton" which is to be distinguished from the
 ``D4 particle" discussed above. The origin  of this term is due to its similarity
   in the 4D Euclidean space-time  to the canonical instanton
(or caloron at $T\neq 0$). This object
tends to condense below the phase transition and is well defined above the
transition point, similar to the instanton. It carries a nontrivial $\theta$ dependence
and the corresponding
contribution to the action from the single "D4 instanton" looks as follows
\begin{equation}
\delta S \propto \theta \int Tr F\wedge F \propto C_1
\end{equation}
Due to the background flux $dC_3$ through $S^4$ one has
\begin{equation}
\label{D-instanton}
N_c \int d x_4 C_1
\end{equation}
term in the action implying that the D4-instanton worldvolume is populated by
N D0 instantons which provide the topological charge N in D4 worldvolume theory.
Due to the unbroken electric $Z_N$ in the confinement phase one could expect
the N-tuples of D4-instantons  to exist. The nature of large factor $N_c$ in  both equations
(\ref{D-instanton}) and  (\ref{D-particle}) is one and the same, namely  the background flux $dC_3$ through $S^4$
which is proportional to $N_c$.
However, the physical interpretation for these two cases is quite different: in the first case
it is the mass of the particle which is $\sim N_c$, while in the second case   it is the action  of  $N_c$
different instantons accompanied by  4$N_c$ zero modes each.
It is tempting to identify this D4- instanton with the configuration
consisting of $N_c$ different calorons with maximally nontrivial holonomies\footnote{ Instanton at
$T\neq 0$ becomes a caloron with a generically  nontrivial holonomy\cite{vanbaal}.}.
As is known, exactly the configuration consisting $N_c$ different calorons provides an infrared finite contribution
to the partition function \cite{Diakonov:2007nv}.  \\

{\it \underline{D4-$\bar{D4}$ U-shaped pair}.}\\
One can also consider the U-shaped D4 brane extended along the radial coordinate.
If it wraps $S^4$ it is an instanton-like object which however does not carry any $\theta$ dependence.
Since there is in fact a  connected $D4-\bar{D4}$ pair one could say that such
objects amount to the chiral symmetry breaking "at a point".
On the other hand one can consider U-shaped D4 brane localized at $S^4$.
Such object is a space-time filling brane which provides a kind of
homogeneous "chiral symmetry breaking" in pure YM theory.



\subsection{D6 branes}
{\it \underline{D6  string
}.}\\
Let us turn to D6 branes. Consider first the pure YM case at large $N_c$.
If D6 is wrapped around $S^4 \times x_4$ it behaves as the string
in the space-time whose tension is defined by the scale of $S^4$.
Due to the wrapping around $S^4$ the string carries a
CS term  generated on its worldsheet from the CS term
\begin{equation}
\int d^7x C_3\wedge F\wedge F
\end{equation}
which reads as $N_c\int d^3 x A\wedge F$. In the
confinement phase   there is a nontrivial holonomy along $x_4$ represented by
$W(R)$, see (\ref{W(R)}),  hence the effective 2d $\theta$ term on the D6 string is induced.
Note that the induced $\theta$ term on t he D6 string is proportional
to $N_c$,  somewhat similar to the situation discussed in \cite{auzzi}
for the magnetic strings in $N=1^(*)$ theory.

Due to the wrapping  around $x_4$ it carries intrinsic
$\theta$ dependent term
\begin{equation}
\int d^7 x C_1\wedge F\wedge F\wedge F
\end{equation}
To provide the $\theta$ dependent contribution from this term
the topological Chern number $c_3(F)$ should be nontrivial.
It follows from the D2 branes on the D6 worldvolume.
Hence we see that the interesting "composite"  D6-D2 string
has $\theta$ dependent contributions from the both components.
Such  D6 strings according to our
standard arguments    are individually unstable and large number of them have  tendency  to condense at small temperatures $T<T_c$.
The object  is well defined
at large temperatures, $T>T_c$ and has finite tension. The interpretation of such kind of objects is far from obvious,
and the role they play in physics is also unclear at the moment.

 One can also consider the U-shaped $D6-\bar{D6}$ pair
 in the confinement phase corresponding to the string with the finite
 tension. Due to the connectness of D6
 brane it localizes the chiral symmetry breaking on its worldvolume.\\

{\it \underline{D6  domain wall
}.}\\
If D6 brane   wraps around $\tau $ it behaves as the domain wall
which is a source of the corresponding RR-form. Such a configuration has been interpreted
in ref. \cite{wittenflux} as the domain wall which separates
different metastable vacua  known to exist in gluodynamics at large $N_c$.
Its worldvolume theory on the domain wall involves the conventional CS term and has no
$\theta$ dependent term.

\section{Composite defects}

In this Section we present a few examples of composite defects
which exhibit interesting features. Some of them may play a crucial role
in understanding of the dynamics.\\

{\it \underline{D0-D2
}.}\\
First, we want to address the following question: what happens to  instantons
 in the confined phase.
  Naively, one could think that as the metric takes the geometry of a cigar
 in $(x_4, u)$ plane at $T<T_c$ the system becomes unstable, and therefore, there is no subject
 for the discussions as instantons simply disappear from the system. However, as we discussed before,
 one should speak about effectively zero action for formation of such kind of objects.
 Therefore, numerous number of these objects can emerge  in the system without any suppression.
In fact, in \cite{Parnachev:2008fy} it has been argued  that this is precisely what is happening when
 one crosses the phase transition line from above.

Now, in order to investigate  what kind of objects may emerge when the phase transition is crossed from above,
we add the D2 domain walls (discussed in section IIIB) localized at some points along
$x_4$ coordinate. In this construction  an object with a fractional topological charge $1/N_c$ may emerge.
Indeed, one can follow the construction of ref. \cite{Davies:1999uw} for SUSY case
when  $N_c$ D2 branes  located symmetrically split the instanton into $N_c$
constituents stretched between pairs of
domain walls. Each constituent has fractional instanton number $1/N_c$  as well as the
fractional monopole number and has no reason to condense.
In our system we have precisely appropriate D2 branes which are needed for this construction.
These monopoles are instantons in the 3d gauge theory on the
D2 worldvolume theory which involves the  scalar corresponding to the position
of D2 branes  on the cigar. As we mention previously in Section IIIA these magnetic monopoles
may play an important role in the region close to the phase transition $0 < |T-T_c|/T_c\leq 1/N_c$.\\

{\it \underline{D6-D4
}.}\\
There are several configurations involving composite D6-D4 defects. The first one to be
mentioned is the combination "D6 string-D4 instanton". Since the D4 instanton shares all coordinates
with D6 string it is melted into the flux on the string of  constant "electric" field.
Another example of the melting concerns the combination "D6 domain wall-D4 particle"
when the D4-particle delocalizes on the domain wall into the flux. It is interesting to note
that since in the holographic QCD the D4 particle is identified with the baryon upon melting
we get a domain wall with the baryonic density. The defect is well defined in the
confinement phase.

 Another interesting possibility concerns the combination "D6 domain wall-D4 instanton".
 In this case a fractional D4 instanton can emerge if there are several domain walls
 localized at different positions at $x_4$ and D4 instanton can be stretched between a
 pair of domain walls in the $x_4$ direction yielding a monopole-like objects on the
 domain-wall  worldvolume.

{\it \underline{D2-D4
}.}\\
An interesting situation happens if we consider the configuration
of D2- string and D4-particle localized in $x_4$ . In this case the magnetic D2
string can be stretched between two D-particles and therefore
does not wrap the $x_4$ circle. Hence it has finite tension equal to the distance
$\delta x_4$ between two D4-particles and does not condense in the
confinement phase. This configuration carries fractional topological
charge and is $\theta$ dependent. The D4-particles acquire  magnetic
charges where the flux of the magnetic string ends on.

In the case of D2-domain wall and D4-instanton opposite situation
can happen. The D4-instanton worldline can be split between
two D2-domain walls localized at different $x_4$ coordinates.
Such configuration carries fractional topological charge as well.\\

{\it \underline{D2-D6
}.}\\
The simplest configuration of such type is " D6 string- D2 string". That
is we have composite string object with additional "charge" since the
D2 induces  instanton-like charge on the D6-brane worldvolume.
Such a composite string is unstable in the confinement phase and well defined in the
deconfinement phase. Another combination "D6 domain wall- D2 string" can be stable in the
confinement phase if there are several domain walls localized at different
positions in $x_4$. The same can be said about the configuration of
" D2 domain wall-D6 string" with several D2 walls.

\section{Defects in holographic QCD }
In this Section we discuss defects in the holographic QCD when matter fields are included
by adding $N_f$ U-shaped D8 branes \cite{ss}  in the probe approximation. This
case can be considered as a particular example of the composite defects
in pure YM theory.
To study defects in holographic QCD we shall add some additional probe branes of  different
dimensions. In what follows we mention   only  a few effects which are specific for theory with the
fundamental matter.

First, we can add U-shaped D6 branes
parallel to the D8 branes. From the 4d viewpoint they are unstable strings.
Indeed the D6 branes share all worldsheet coordinates with the D8 branes,
hence there are tachyonic modes in the spectrum of D8-D6 open strings
and the D6 brane melts in the D8 worldvolume yielding a flux of the flavor
gauge  field. Since the D6 branes get delocalized they do not provide
string-like localization of the chiral symmetry breaking.

Another interesting example concerns the D4 instanton in holographic QCD
extended along $x_4$ or radial coordinate.  To estimate its contribution
into the partition function remind that there is CS term on the D8 brane
worldvolume multiplied by $N_c$. Since U-shaped D4 instanton shares all coordinates
with the D8 branes it induces a nontrivial contribution into the CS part of the action which
reads as follows
\begin{equation}
\delta S_{CS}=N_c\int du A_u \int d^4 TrF\wedge F
\end{equation}
where we assume that $U(1)_{A}$ field $A_u$ is space-time independent. It can be interpreted as
the constant mode of $\eta'$ meson since in the SS model it is identified as $\int du A_{u}(u,x)$.
Note that  that the constant mode of $\eta'$ mode can be thought of as the effective
$\theta$ term hence the total contribution reads as $\theta_{eff}N_c k$ where k- is
the instanton number in the flavor gauge theory. Such U-shaped D4 instanton is well defined
at any temperature since it does not wrap around any compact coordinate and one could speak
about the point-like contribution to the chiral symmetry breaking.

One can also discuss fractional flavor D4 instantons.
To this aim the flavor branes have to be placed at different positions at the dual
temporal circle. This can be achieved by switching on  chemical potentials
which correspond to nontrivial temporal holonomies of the flavor gauge fields.
The eigenvalues of the flavor holonomy around the temporal circle provide the
positions of the D7 branes on the dual circle. Hence
the worldline of the D5 brane on the dual circle can split  and we get $N_f$ fragments of D5 brane stretched
between $N_f$ D7 branes which carry the flavor "magnetic" and fractional instanton charges.
 The system can be described in  QFT terms as a set of $N_f$ "monopoles" with the total instanton charge
$Q_{inst}=1$  and the total monopole charge zero, similar to the canonical caloron  with nontrivial temporal holonomies, see
for reviews \cite{gross,Bruckmann:2004zy} and references therein. Since we are working in the probe approximation
 $N_f<<N_c$  the action on each "monopole" is proportional to $N_C$  and therefore these defects
 are suppressed at large $N_c$.


Another possible phenomenon concerns a peculiar manifestation
of the Myers effect in the holographic QCD. It is known that
the following configuration: two parallel D4 branes + D0 branes
localized on the D4 branes + D0-D0 open strings, is unstable and
decays into the so-called dyonic instanton \cite{dyonic}. That is this configuration
of instantons decays into the circular D2 brane stretched between parallel
D4 branes with   electric and topological charges as well
as  angular momentum which stabilizes the system.
On the D4 worldvolume one gets a magnetically charged ring.

In our context, we can consider the set of D8 and D4 branes representing the
baryons in the space-time \cite{baryons}. We assume that the worldvolume of the D4 brane
is $S^4\times \tau$, that is both type of branes are localized
at the $x_4$ coordinate. We could assume that the D8 branes are
generically localized at different values $x_{4k}$ where $k+1,\dots N_f$.
and the "baryonic" D4 branes are localized on the D8 branes. Consider
two baryonic D4 branes and connect them by an open string which
carries in the spectrum of excitations the gauge boson of the flavor
group- $\rho$ meson. Similar to the D4-D0-F1 case we expect that
the D4 brane is expanded into a D6 brane with baryonic density
and the F1 string is melted into the isotopic charge amounted
from the initial $\rho$-meson. Hence finally we could expect
that the configuration with the baryonic charge larger than one
gets delocalized into the "baryonic ring" carrying the isotopic charge.
Note that it is known in the Skyrme model that the B=2 state has a torus
like ground-state geometry \cite{stern}, in agreement with our arguments above.
Note that such a configuration does not involve  wrapping around $x_4$,
that is it is well defined  at $T<T_c$.

\section{Conclusions}

In this note we  demonstrated that the holographic
description of the pure YM or QCD-like theories implies
existence, at least classically of a plenty of  defects
of different codimensions. We have tried to argue
that their existence is insensitive to  details of the
metric. However the analysis is certainly oversimplified  and
we have not aimed at deriving the defects
properties in detail. We have seen that various types of strings
and domain walls and some more exotic composite objects are emerging.

A few claims seem to be quite generic. If the defect tends to
condense (have small tension vanishing classically) at small
temperatures it becomes tensionful and well defined above the
phase transition. On the other hand, all the defects apart from the S-branes
tend to loose one Euclidean dimension above the critical temperature.
The $\theta$ dependence of the defect's worldvolume actions
is present in the "condensing brane" below the phase transition
and can be defined on such defects above the transition point
as well on a single defect. This is  an analogue
of the instanton solution of the U(1) problem via the Witten-Veneziano
mechanism below the transition point and via the t'Hooft mechanism above
this point. Our analysis  implies that a similar change
of mechanisms happens for defects of  different dimensions as well.

The theory above the phase transition almost immediately
becomes three-dimensional
because of the cigar-type instability. This fits well with the lattice studies.

In particular, in holographic description we identified a few very interesting objects such as
  heavy buckyballs (D4 particles) whose  masses scale as $\sim N_c$ (see section IIIC)  or
 D4 instantons whose action scales as $\sim N_c$.
While such objects have been discussed previously in the literature
within QFT, their future study using the holographic description
may shed a new light on their nature.
Another interesting example deserves to be mentioned is the holographic description
of the objects with fractional topological and magnetic charges (section IV).
Such objects  have been  discussed within QFT in late 70s.
Future study of these objects may provide  with a key to understand the  QCD vacuum structure and the nature of the phase transition.

There are many questions we have only touched upon. In particular, it would be
highly interesting to understand better the role of the second order
parameter "dual" to the Polyakov loop and the
duality between the corresponding pairs of domain walls which gets interchanged
at the phase transition. Our analysis suggests that
it is probably reasonable to discuss the chiral symmetry breaking
in pure YM theory induced by U-shaped defects. At first glance it might look
strange, but " chiral symmetry" of the gauge boson can be defined similar to fermions.
The order parameter for such a "chiral symmetry breaking" could be  similar
to the one recently discussed in \cite{aku}. We did not discuss the defects
involving NS5 branes since it is more natural to discuss such issue upon
lifting of IIA setup to the M-theory. In that case the corresponding
defects involve M2, M5, KK particles and KK monopoles.

\acknowledgments
 We   thank The Galileo Galilei Institute for Theoretical Physics for
the organization of the workshop ``Non-perturbative methods in strongly coupled
gauge theories'' where this  projected  was initiated.

  ARZ
was supported, in part, by the Natural Sciences and Engineering
Research Council of Canada. The work  of A.G. was supported in part by grants
INTAS-1000008-7865 and PICS- 07-0292165. V.I.Z. would like to thank the Laboratoire de
Mathematiques et Physique Theorique at UNiversi\'e Francois Rabelais, Tours, France for kind
hospitality during the final stages of working on the project.


\begin{thebibliography}{1234567}

\bibitem{wittenterm}
E.~Witten,
  ``Anti-de Sitter space, thermal phase transition, and confinement in  gauge
  theories,''
  Adv.\ Theor.\ Math.\ Phys.\  {\bf 2}, 505 (1998)
  [arXiv:hep-th/9803131].


\bibitem{ss}
T. Sakai and S. Sugimoto, "Low energy hadron physics in holographic
QCD," Prog. Theor. Phys. 113, 843 (2005) [arXiv:hep-th/0412141];
"More on a holographic dual of QCD," Prog. Theor. Phys. 114, 1083
(2005) [arXiv:hep-th/0507073

\bibitem{Bergman:2006xn}
  O.~Bergman and G.~Lifschytz,
 ``Holographic U(1)A and string creation,''
  JHEP {\bf 0704}, 043 (2007)
  [arXiv:hep-th/0612289].
 \bibitem{gz}
A. Gorsky, V. Zakharov, {\it ``Magnetic strings in Lattice QCD as
Nonabelian Vortices''},   [arXiv:0707.1284 [hep-th]].


\bibitem{wittenflux}
E. Witten,  `` Theta dependence in the large N limit of
four-dimensional gauge theories'',  Phys. Rev. Lett. {\bf  81}
 2862 (1998) [arXiv:hep-th/9807109].

\bibitem{Lin:2008vv}
  F.~L.~Lin and S.~Y.~Wu,
  ``Holographic QCD with Topologically Charged Domain-Wall/Membranes,''
  JHEP {\bf 0809}, 046 (2008)
  [arXiv:0805.2933 [hep-th]].

\bibitem{Hong:2008nh}
  D.~K.~Hong, K.~M.~Lee, C.~Park and H.~U.~Yee,
  ``Holographic Monopole Catalysis of Baryon Decay,''
  JHEP {\bf 0808}, 018 (2008)
  [arXiv:0804.1326 [hep-th]].




\bibitem{wittenah}
  O.~Aharony and E.~Witten,
  ``Anti-de Sitter space and the center of the gauge group,''
  JHEP {\bf 9811}, 018 (1998)
  [arXiv:hep-th/9807205].

\bibitem{armoni}
  A.~Armoni, S.~P.~Kumar and J.~M.~Ridgway,
  ``Z(N) Domain walls in hot N=4 SYM at weak and strong coupling,''
  arXiv:0812.0773 [hep-th].

\bibitem{yee}
  H.~U.~Yee,
  ``Fate of Z(N) domain wall in hot holographic QCD,''
  arXiv:0901.0705 [hep-th].



\bibitem{douglas}
 M. R. Douglas, "Branes within branes," arXiv:hep-th/9512077.

\bibitem{gava}
 E.~Gava, K.~S.~Narain and M.~H.~Sarmadi,
  ``On the bound states of p- and (p+2)-branes,''
  Nucl.\ Phys.\  B {\bf 504}, 214 (1997)
  [arXiv:hep-th/9704006].

 \bibitem{dia}
    D.~E.~Diaconescu,
  ``D-branes, monopoles and Nahm equations,''
  Nucl.\ Phys.\  B {\bf 503}, 220 (1997)
  [arXiv:hep-th/9608163].


\bibitem{myers}
  R.~C.~Myers,
  ``Dielectric-branes,''
  JHEP {\bf 9912}, 022 (1999)
  [arXiv:hep-th/9910053].




\bibitem{Toublan:2005tn}
  D.~Toublan and A.~R.~Zhitnitsky,
  ``Confinement - deconfinement phase transition at nonzero chemical
  potential,''
  Phys.\ Rev.\  D {\bf 73}, 034009 (2006)
  [arXiv:hep-ph/0503256].




\bibitem{Parnachev:2008fy}
  A.~Parnachev and A.~Zhitnitsky,
  ``Phase Transitions, theta Behavior and Instantons in QCD and its Holographic
  Model,''
  Phys.\ Rev.\  D {\bf 78}, 125002 (2008)
  arXiv:0806.1736 [hep-ph].

\bibitem{Zhitnitsky:2008ha}
  A.~R.~Zhitnitsky,
  ``Confinement- Deconfinement Phase Transition in Hot and Dense QCD at Large N,''
  Nucl.\ Phys.\  A {\bf 813}, 279 (2008)
  [arXiv:0808.1447 [hep-ph]].


\bibitem{Alles:1996nm}
  B.~Alles, M.~D'Elia and A.~Di Giacomo,
  ``Topological susceptibility at zero and finite T in SU(3) Yang-Mills
  theory,''
  Nucl.\ Phys.\  B {\bf 494}, 281 (1997)
  [Erratum-ibid.\  B {\bf 679}, 397 (2004)]
  [arXiv:hep-lat/9605013].

\bibitem{Lucini:2003zr}
  B.~Lucini, M.~Teper and U.~Wenger,
  ``The high temperature phase transition in SU(N) gauge theories,''
  JHEP {\bf 0401}, 061 (2004)
  [arXiv:hep-lat/0307017].

\bibitem{Lucini:2004yh}
  B.~Lucini, M.~Teper and U.~Wenger,
  ``Topology of SU(N) gauge theories at T approx. 0 and T approx. T(c),''
  Nucl.\ Phys.\  B {\bf 715}, 461 (2005)
  [arXiv:hep-lat/0401028].

\bibitem{Del Debbio:2004rw}
  L.~Del Debbio, H.~Panagopoulos and E.~Vicari,
  ``Topological susceptibility of SU(N) gauge theories at finite
  temperature,''
  JHEP {\bf 0409}, 028 (2004)
  [arXiv:hep-th/0407068].

\bibitem{Lucini:2005vg}
  B.~Lucini, M.~Teper and U.~Wenger,
  ``Properties of the deconfining phase transition in SU(N) gauge theories,''
  JHEP {\bf 0502}, 033 (2005)
  [arXiv:hep-lat/0502003].

\bibitem{Vicari:2008jw}
  E.~Vicari and H.~Panagopoulos,
  ``Theta dependence of SU(N) gauge theories in the presence of a topological
  term,''
  arXiv:0803.1593 [hep-th].

\bibitem{kutasov1}
E. Antonyan, J.A. Harvey and D. Kutasov,
``The Gross-Neveu Model from String Theory'',
 Nucl.Phys. {\bf B776} (2007) 93, [arXiv:hep-th/06081].

\bibitem{vanbaal}  T.C. Kraan and P. van Baal,
Nucl. Phys. {\bf B533} (1998) 627; Phys. Lett.{\bf  B428}  (1998) 268


 \bibitem{sign}
  M.~N.~Chernodub, K.~Ishiguro, A.~Nakamura, T.~Sekido, T.~Suzuki and V.~I.~Zakharov,
  ``Topological defects and equation of state of gluon plasma,''
  PoS {\bf LAT2007}, 174 (2007)
  [arXiv:0710.2547 [hep-lat]].

\bibitem{langfeld}
M. Engelhardt, K. Langfeld, H. Reinhard and O. Tennert,
`` Deconfinement in SU(2) Yang-Mills theory as a center vortex percolation transition'',
 Phys. Rev. {\bf D61} (2000) 054504,
 [hep-lat/9904004]\\
M.N. Chernodub, A. Nakamura and V.I. Zakharov,
`` Abelian monopoles and center vortices in Yang-Mills plasma'',
  arXiv:0812.4633 [hep-ph]

\bibitem{Csorgo:2001sq}
  T.~Csorgo, M.~Gyulassy and D.~Kharzeev,
  ``Buckyballs and gluon junction networks on the femtometer scale,''
  J.\ Phys.\ G {\bf 30}, L17 (2004)
  [arXiv:hep-ph/0112066].

  \bibitem{fullerenes}H.W.Kroto, R.E.Smalley and R.F.Curl, Nature {\bf 318}  (1985) 165

\bibitem{Diakonov:2007nv}
  D.~Diakonov and V.~Petrov,
  ``Confining ensemble of dyons,''
  Phys.\ Rev.\  D {\bf 76}, 056001 (2007)
  [arXiv:0704.3181 [hep-th]].



\bibitem{auzzi}
  R.~Auzzi and S.~Prem Kumar,
  ``Non-Abelian k-Vortex Dynamics in $N=1^*$ theory and its Gravity Dual,''
  JHEP {\bf 0812}, 077 (2008)
  [arXiv:0810.3201 [hep-th]].

\bibitem{Davies:1999uw}
  N.~M.~Davies, T.~J.~Hollowood, V.~V.~Khoze and M.~P.~Mattis,
  ``Gluino condensate and magnetic monopoles in supersymmetric  gluodynamics,''
  Nucl.\ Phys.\  B {\bf 559}, 123 (1999)
  [arXiv:hep-th/9905015].

\bibitem{gross}
  D.~J.~Gross, R.~D.~Pisarski and L.~G.~Yaffe,
  ``QCD And Instantons At Finite Temperature,''
  Rev.\ Mod.\ Phys.\  {\bf 53}, 43 (1981).

\bibitem{Bruckmann:2004zy}
  F.~Bruckmann, E.~M.~Ilgenfritz, B.~V.~Martemyanov, M.~Muller-Preussker, D.~Nogradi, D.~Peschka and P.~van Baal,
  ``Calorons with non-trivial holonomy on and off the lattice,''
  Nucl.\ Phys.\ Proc.\ Suppl.\  {\bf 140}, 635 (2005)
  [arXiv:hep-lat/0408036].
\bibitem{baryon}
 E. Witten, "Baryons and branes in anti de
Sitter space", JHEP 9807, 006 (1998) [arXiv:hep-th/9805112]


 \bibitem{dyonic}
  P.~K.~Townsend,
  ``Field theory supertubes,''
  Comptes Rendus Physique {\bf 6}, 271 (2005)
  [arXiv:hep-th/0411206].\\
  H.~Y.~Chen, M.~Eto and K.~Hashimoto,
  ``The shape of instantons: Cross-section of supertubes and dyonic
  instantons,''
  JHEP {\bf 0701}, 017 (2007)
  [arXiv:hep-th/0609142].


 \bibitem{stern}
  V.~B.~Kopeliovich and B.~E.~Stern,
  ``Exotic Skyrmions,''
  JETP Lett.\  {\bf 45}, 203 (1987)
  [Pisma Zh.\ Eksp.\ Teor.\ Fiz.\  {\bf 45}, 165 (1987)].


\bibitem{aku}
  O.~Aharony and D.~Kutasov,
  ``Holographic Duals of Long Open Strings,''
  Phys.\ Rev.\  D {\bf 78}, 026005 (2008)
  [arXiv:0803.3547 [hep-th]].




\exclude{

\bibitem{Chernodub:2006gu}
  M.~N.~Chernodub and V.~I.~Zakharov,
  ``Magnetic component of Yang-Mills plasma,''
  Phys.\ Rev.\ Lett.\  {\bf 98}, 082002 (2007)
  [arXiv:hep-ph/0611228].


\bibitem{ooguri}
 E. Witten, Baryons and branes in anti de
Sitter space, JHEP 9807, 006 (1998) [arXiv:hep-th/9805112];\\
 D. J. Gross and H.
Ooguri, Aspects of large N gauge theory dynamics as seen by string
theory, Phys. Rev. D 58, 106002 (1998) [arXiv:hep-th/9805129].



\bibitem{Hata:2007mb}
  H.~Hata, T.~Sakai, S.~Sugimoto and S.~Yamato,
  ``Baryons from instantons in holographic QCD,''
  arXiv:hep-th/0701280.

\bibitem{sugimoto}
K. Hashimoto, T. Sakai, Sh. Sugimoto, {\it``Holographic Baryons :
Static Properties and Form Factors from Gauge/String Duality''},
 [arXiv:0806.3122 [hep-th]].





\bibitem{Fateev} V.Fateev et al, Nucl. Phys. {\bf B154}
(1979) 1;
B.Berg and M.Luscher, Commun.Math.Phys. {\bf 69}(1979) 57.
\bibitem{Belavin} A. Belavin et al, Phys. Lett. {\bf 83B} (1979) 317.



\bibitem{JZ} S. Jaimungal and A.R. Zhitnitsky,  [hep-ph / 9904377],  [hep-ph / 9905540],
unpublished.  \\   A.~R.~Zhitnitsky,
  ``Confinement - deconfinement phase transition and fractional instanton
  quarks in dense matter,''
  arXiv:hep-ph/0601057.
 }








\end{thebibliography}
 \end{document}